\documentclass[a4paper,preprint,aps,prd]{revtex4-1}
\usepackage{amssymb} 
\usepackage{amsmath} 
\usepackage{textcomp}
\usepackage[pdftex]{graphicx} 
\usepackage{subfigure} 
\usepackage{caption} 
\usepackage{color}
\usepackage{hyperref}

\begin{document}

\title{Prospects for the detection of the Diffuse Supernova Neutrino\\ Background with the experiments SK-Gd and JUNO}

\author{Yu-Feng Li}
\affiliation{Institute of High Energy Physics, Chinese Academy of Sciences and School of Physical Sciences, University of Chinese Academy of Sciences, Beijing 100049, China;
 liyufeng@ihep.ac.cn}
\author{Mark Vagins}
\affiliation{Kavli Institute for the Physics and Mathematics of the Universe (WPI),\\ The University of
 Tokyo Institutes for Advanced Study, University of Tokyo, Kashiwa, Chiba 277-8583, Japan; mark.vagins@ipmu.jp}
\affiliation{Department of Physics and Astronomy, University of California, Irvine, Irvine, CA 92697-4575, USA}
\author{Michael Wurm}
\affiliation{Institute of Physics and Excellence Cluster PRISMA$^+$, Johannes Gutenberg-Universit\"at Mainz, Mainz, Germany; michael.wurm@uni-mainz.de \vspace{5mm}}

\date{\today}

\begin{abstract}
\noindent The advent of gadolinium-loaded Super-Kamiokande (SK-Gd) and of the soon-to-start JUNO liquid scintillator detector marks a substantial improvement in the global sensitivity for the Diffuse Supernova Neutrino Background (DSNB). The present article reviews the detector properties most relevant for the DSNB searches in both experiments and estimates the expected signal and background levels. Based on these inputs, we evaluate the sensitivity of both experiments individually and combined. Using a simplified statistical approach, we find that both SK-Gd and JUNO have the potential to reach $>$3$\sigma$ evidence of the DSNB signal within 10 years of measurement. The combined results are likely to enable a $5\sigma$ discovery of the DSNB signal within the next decade.
\end{abstract}

\maketitle

\section{Introduction}

 Core-collapse Supernovae (SNe) count among the brightest sources of low energy neutrinos ($E_\nu\lesssim 50$\,MeV). A supernova occurring within the Milky Way will cause an intense burst of events in currently running neutrino detectors. The signal will encode details of the astrophysics of the explosion superimposed with the effects of neutrino properties and oscillations (for a comprehensive review, see e.g., Ref.\ \cite{Mirizzi:2015eza}). However, even compared to the decades of operation of large-volume neutrino observatories, galactic SNe are rare. This makes the Diffuse Supernova Neutrino Background (DSNB), i.e.,~the faint but constant flux of neutrinos emitted by core-collapse SNe on cosmological distances, an attractive research objective \cite{Ando:2004hc,Beacom:2010kk,Vissani:2011kx,Lunardini:2012ne,Nakazato:2015rya,Horiuchi:2017qja,Priya:2017bmm,Moller:2018kpn,Riya:2020wpw,Kresse:2020nto}. A first measurement of the DSNB has the potential to provide valuable information on the redshift-dependent SN rate as well as on the average and variability of the SN neutrino spectrum.

Given the minute expected flux of ${\cal O}(10^2)$ per cm$^2$s and red-shifted energy of DSNB neutrinos and anti\-neutrinos of all flavors, an experimental observation has proven to be very challenging. Detector target masses on the order of $\sim$10 kilotons are required to obtain one signal event per year. The current best upper limit on the DSNB's $\bar\nu_e$ flux component is held by the Super-Kamiokande (Super-K, SK) water Cherenkov experiment at 2.7\,cm$^{-2}$s$^{-1}$ above 17.3\,MeV \cite{Super-Kamiokande:2021jaq}. This result is already cutting into the parameter range predicted by current DSNB models (e.g., \,\cite{Kresse:2020nto}). 

During the next decade, a first detection of the long-sought DSNB signal is finally coming within reach. The two neutrino observatories most likely to achieve first evidence ($3\sigma$) of the DSNB signal are Super-Kamiokande and JUNO. In 2020, the Super-Kamiokande collaboration has performed an upgrade of the detector by dissolving gadolinium salt in the water target. This greatly enhances neutron detection capabilities \cite{Vagins:2005ii,Vagins:2007zz,Sekiya:2016xji}, leading to a significant improvement in the efficiency and background rejection for the Inverse Beta Decay (IBD) detection channel and thus the $\bar\nu_e$ component of the DSNB. Data taking in the new SK-Gd configuration commenced in August 2020. In parallel, the JUNO liquid scintillator (LS) experiment in southern China is entering its construction phase \cite{JUNO:2015zny}. With first data expected in 2023, JUNO will acquire IBDs at a rate only slightly lower than SK-Gd, relying on the intrinsic neutron tag and pulse-shape discrimination (PSD) capabilities of liquid scintillator \cite{JUNO:2015zny,JUNO:2021vlw}.

We would like to note that beyond the operational SK-Gd and the soon-to-be operational JUNO, there are a number of other experiments on the horizon with varying degrees of sensitivity to the DSNB. In particular, Hyper-Kamiokande, which is currently under construction, will directly continue the search of SK from $\sim$2027 using eight times SK’s fiducial volume~\cite{Abe:2011ts}.  This is briefly discussed in Section~\ref{sec:hk}.  Large noble-liquid detectors, while challenged by expected low signal event rates and as-yet undetermined backgrounds, could in principle provide sensitivity for other neutrino flavors (DUNE/liquid argon for $\nu_e$, DARWIN/liquid xenon for $\nu_{\mu,\tau}$ flavors), while conceptual hybrid Cherenkov-scintillation detectors such as Theia, if someday realized, could feature enhanced detection efficiencies for $\bar\nu_e$'s~\cite{Abi:2018dnh, Suliga:2021hek, Wei:2016vjd, Sawatzki:2020mpb}.

The present article aims to review the DSNB detection potential of the two experiments. Based on the relatively simple model of the DSNB flux and spectrum presented in Section~\ref{sec:dsnb1}, we discuss the signal and background rates expected for SK-Gd and JUNO (Sections~\ref{sec:sk} and \ref{sec:juno}). Based on these numbers, Section~\ref{sec:sensitivity} tracks the signal rates and sensitivities of both experiments as a function of their respective measuring times. Since both experiments can hope to gain first $3\sigma$-evidence of the DSNB signal within the next decade, a $5\sigma$-observation may be achieved by a combination of their results over a similar time scale.
 
\section{Signal of the Diffuse Supernova Neutrino Background}
\label{sec:dsnb1}

The DSNB flux and spectrum results from a superposition of the neutrino bursts from core-collapse SNe happening on cosmic distance scales. Given the large numbers and distances to the parent SNe, the resulting DNSB flux is of the order of $10^2$ per cm$^2$s and nearly isotropic. The effective energy spectrum represents an average of the entire population of stellar core collapses from a wide range of progenitor stars, including failed explosions that lead to the formation of a Black Hole (BH). Spectral contributions from far-out SNe are substantially red-shifted. Hence, the signal range detectable in SK-Gd and JUNO (above $\sim$10\,MeV, see below) is dominated by relatively close-by SNe up to red-shifts $z \approx 1$ (see, e.g., Ref.~\cite{Ando:2004hc}).

The expectation for the differential electron antineutrino flux of the DSNB is given by the integral
\begin{equation} \label{eq:diffflux}
    \frac{d\Phi(E_\nu)}{dE_\nu}=\ \frac{c}{H_{0}}\int_{0}^{z_{\rm max}} {R_{\rm CC}\left(z\right)\frac{dN_\nu(E'_\nu)}{dE_\nu}\frac{dz}{\sqrt{{\Omega}_{\Lambda}+\Omega_m(1+z)^3}}},
\end{equation} 
where $E_\nu$ ($E'_\nu$) is the (redshifted) neutrino energy, $c$ is the speed of light and $H_0$, $\Omega_\Lambda$, $\Omega_m$ are cosmological parameters (e.g.~\cite{Priya:2017bmm}). $R_{\rm CC}(z)$ is the redshift-dependent rate of core-collapse SNe, whose $z$ dependence is derived from the star formation rate~\cite{Hopkins:2006bw} with the \mbox{following relation}:
\begin{equation}\label{R_SF}
R_\mathrm{CC}(z) = R_\mathrm{CC}(0)\frac{(a+bz)h}{ah[1+(z/c)^d]}\,,
\end{equation}
where $a=0.0170$, $b=0.13$, $c=3.3$, $d=5.3$ and $h = 0.7$ parametrize the $z$-dependence. $R_\mathrm{CC}(0)$ is the present rate of core-collapse SNe and taken as $1.0\times 10^{-4} \mathrm{yr^{-1}\,Mpc^{-3}}$ in the following DSNB \emph{reference model}.

An important choice for the DSNB modeling is the average SN neutrino energy spectrum $ {{dN}/{dE_{\nu}}}$. In accordance with Ref.~\cite{Priya:2017bmm}, we take into account the contributions from both successful and failed SNe:
\begin{equation}\label{Flux_CCSN}
\frac{dN(E_\nu)}{dE_\nu} = (1-f_\mathrm{BH})\frac{dN_\mathrm{SN}(E_\nu)}{dE_\nu}
+ f_\mathrm{BH} \frac{dN_\mathrm{BH}(E_\nu)}{dE_\nu},
\end{equation}
with $f_\mathrm{BH}$ indicating the fraction of black hole (BH) forming core-collapse SNe in the total event sample.

The average energy spectrum for both types of SNe can be parametrized as
\begin{equation}\label{Spectrum_CCSN}
\frac{dN_\nu}{dE_\nu} = \frac{E_\mathrm{total}}{\langle E_\nu \rangle^2}
\frac{(1+\gamma_\alpha)^{1+\gamma_\alpha}}{\Gamma(1+\gamma_\alpha)}
\left( \frac{E_\nu}{\langle E_\nu \rangle} \right)^{\gamma_\alpha} \exp\left( -(1+\gamma_\alpha) \frac{E_\nu}{\langle E_\nu \rangle} \right),
\end{equation}		
where $ E_\mathrm{total} $ is the total energy emitted, $ \langle E_\nu \rangle $ is the average energy of the SN neutrino spectrum, and
\begin{equation}\label{gamma_alpha}
\gamma_\alpha = \frac{\langle  E_\nu^2 \rangle - 2 \langle  E_\nu \rangle^2 }
{\langle  E_\nu \rangle^2 -\langle  E_\nu^2 \rangle}
\end{equation}
describes the spectral deviation from a thermal Fermi-Dirac spectrum (pinching) \cite{Keil:2003}.

Inspired by the current state-of-the-art on DSNB modeling, we choose the following parameters to define our DSNB \emph{reference model}: For successful SNe, we take $ E_\mathrm{total} = 5.0\times 10^{52} \ \mathrm{erg} $, $\gamma_\alpha = 3$ and $\langle  E_\nu \rangle = 15\ \mathrm{MeV}$. For failed SNe, we assume $E_\mathrm{total} = 8.6\times 10^{52} \ \rm{erg} $, $ \langle E_\nu \rangle = 18.72 \ \rm{MeV} $ and $ \langle  E_\nu^2 \rangle = 470.76  $ as in Ref.~\cite{Priya:2017bmm}. For the relative fraction of BH forming SNe, we use $f_\mathrm{BH} = 0.27$ adopted from Refs.~\cite{Horiuchi:2017qja,Priya:2017bmm}.

\begin{figure}[t]	

\includegraphics[width=0.6\textwidth]{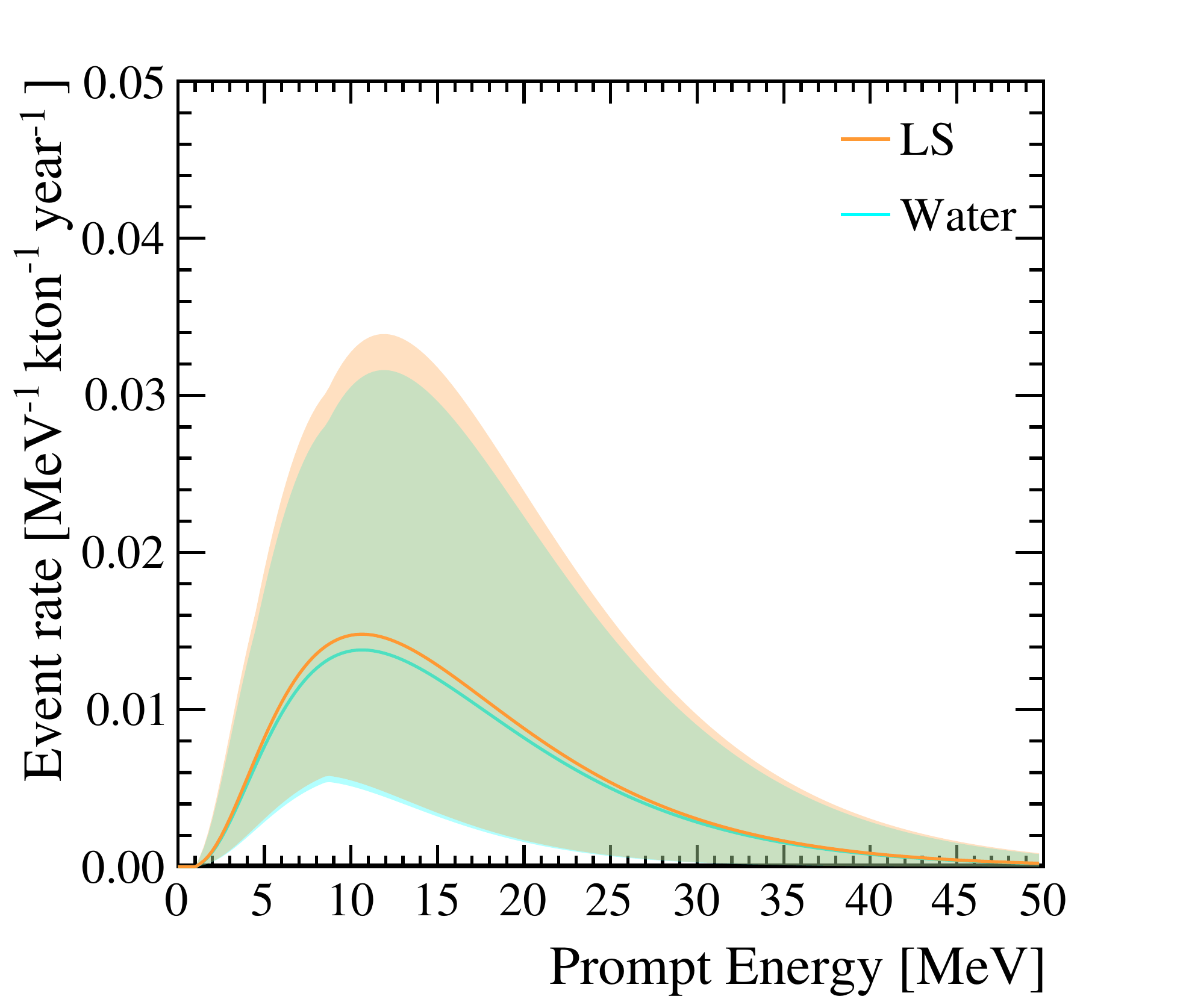}
\caption{The DSNB interaction rates as a function of the prompt energy of the IBD reaction for WC and LS detectors. Shaded areas reflect the impact of the range of parameter predictions listed in Table~\ref{tab:DSNB_model} on the expected rates. \label{fig:ibd_rates}}
\end{figure}

Given that many of the discussed input parameters are not known with great precision, the actual DSNB spectrum might deviate considerably from our DSNB \emph{reference model}. Consequently, we have introduced value ranges for the parameters that have the largest impact on the final DSNB event rate. In particular, we scan $\langle  E_\nu \rangle$ from 12 to 18 MeV, $ f_\mathrm{BH}$ from 0 to 40\%, and
$0.5 \times 10^{-4}\,\mathrm{yr^{-1}\,Mpc^{-3}}\leq R_\mathrm{SN}(0)\leq 2.0 \times 10^{-4}\,\mathrm{yr^{-1}\,Mpc^{-3}} $. The corresponding variability in the signal prediction is indicated by the shaded areas in Figure  \ref{fig:ibd_rates}. The parameters and ranges of the \emph{reference model} are summarized in Table \ref{tab:DSNB_model}. We note that the relatively wide ranges quoted implicitly envelope a wide span of astrophysical observations (e.g.~the soft neutrino spectrum emitted by SN1987A or possible variations in the total explosion energy) and the effects of flavor oscillations on the detected $\bar\nu_e$ spectrum (with a potential for spectral hardening by the admixture of a higher-temperature $\nu_x$ component).

\begin{table}[t] 
\setlength{\tabcolsep}{3.4mm}
\caption{Parameters of the DSNB \emph{reference model} based on current most-likely predictions \cite{Priya:2017bmm}. The parameter ranges adopted  to reflect the uncertainties of these predictions are indicated in brackets. \label{tab:DSNB_model}}
\centering
\begin{tabular}{ccccc}
\hline
\textbf{Parameter}	& \multicolumn{2}{c}{\textbf{Successful SNe}} & \multicolumn{2}{c}{\textbf{Failed SNe}} \\
\hline
Total energy $E_{\rm total}$ [erg] & $5.0\times 10^{52}$ & & $ 8.6\times 10^{52}$ \\
Mean energy $\langle E_\nu \rangle$ [MeV] & 15 & (12 .. 18)&  18.72\\
Relative fraction $f_{\rm BH}$ & 0.73 & $(1-f_{\rm BH})$ & 0.27 & (0 .. 0.4) \\
\hline
Present SN rate $R_{\rm SN}(0)$ & $1.0\times 10^{-4}$ & \multicolumn{2}{l}{$(0.5\,..\,2.0)\times 10^{-4}$} & $\mathrm{yr^{-1}\,Mpc^{-3}}$\\
\hline
\end{tabular}
\end{table}

To obtain the energy-dependent interaction rate $d R/dE_\nu$ of electron antineutrino interactions shown in Figure \ref{fig:ibd_rates}, we evaluate the product
\begin{equation}
      \frac{d R(E_\nu)}{dE_\nu}= \frac{d \Phi(E_\nu)}{dE_\nu}\cdot\sigma_{\rm IBD}(E_\nu)\cdot N_p
\end{equation}
where $\sigma_{\rm IBD}(E_\nu)$ is the IBD cross-section taken from \cite{Strumia:2003zx} and $N_p$ is the number of free protons contained per unit detector mass. Figure \ref{fig:ibd_rates} depicts the interaction rates as function of the prompt energy, i.e.,~the energy of the positron created in the IBD reaction that is experimentally observable. Due to the reaction kinematics, the prompt positron signal nearly preserves the energy information of the initial $\bar\nu_e$. The final-state neutron thermalizes by scattering off hydrogen in the water/scintillator targets within \textmu{}s, and is later on captured either on hydrogen with $\tau_n={\cal O}(200\,\mu{\rm s}$) or considerably faster in case of gadolinium-loading. Detecting the gamma ray(s) from the delayed captures will be the key ingredient for a successful DSNB detection (see below).

\section{Super-Kamiokande with Gadolinium-Doping (Sk-Gd)}
\label{sec:sk}


\subsection{A Brief History of Super-Kamiokande}
\begin{figure}[t]	
\includegraphics[width=0.75\textwidth]{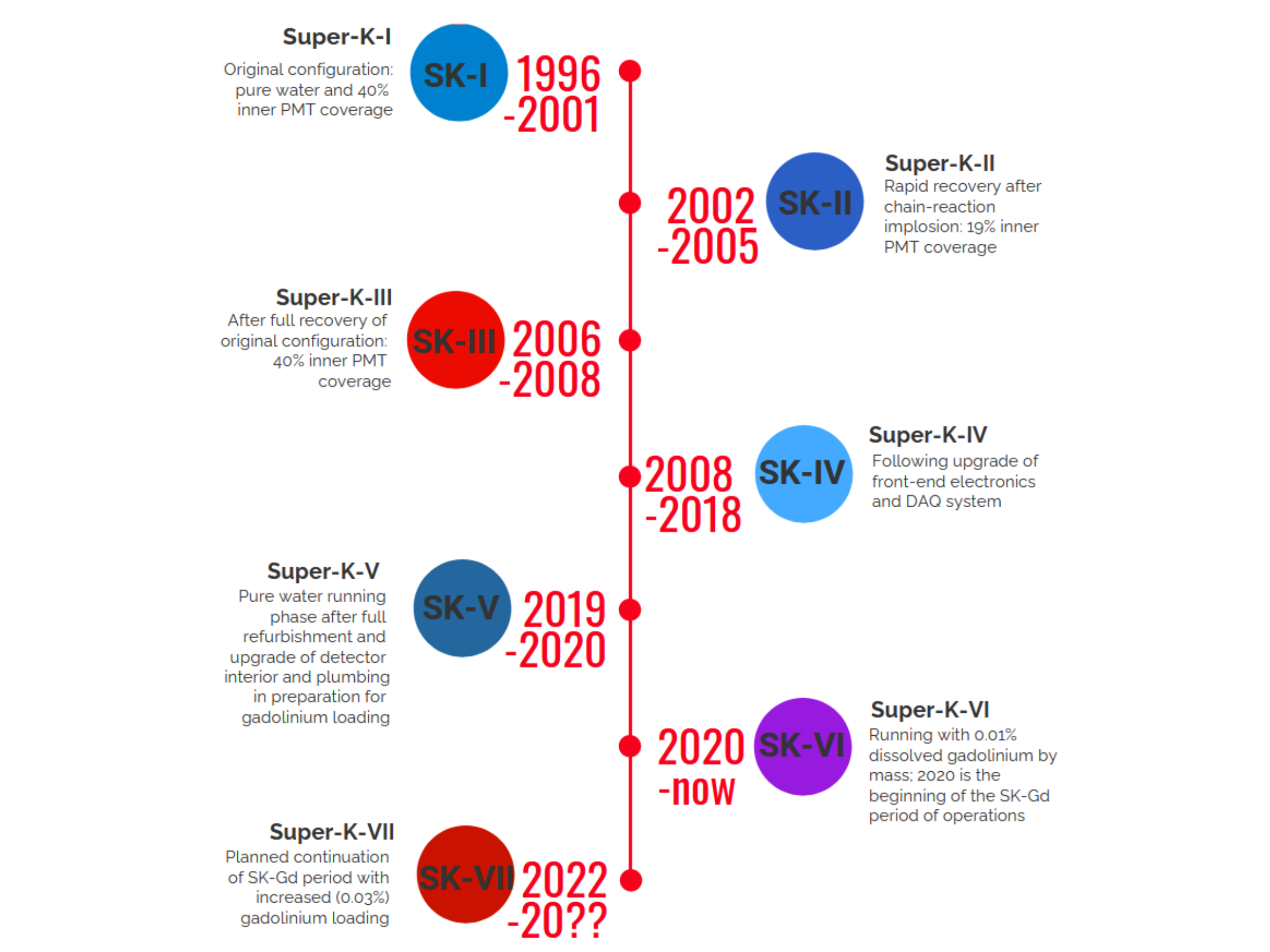} 
\caption{History of Super-Kamiokande's operational phases. For SK-I through
SK-V the detector was filled with ultrapure water. From SK-VI in 2020 onward 
the detector's water has dissolved gadolinium in it to enhance neutron visibility.  \label{SK_hist}}
\end{figure}

Since the start of data taking on 1st April 1996, the 
Super-Kamiokande experiment has 
spent the last quarter century conducting ground-breaking studies of neutrinos from the 
Earth's atmosphere~\cite{Super-Kamiokande:1998kpq}, the Sun~\cite{Super-Kamiokande:2001ljr,Super-Kamiokande:2013mie}, and 
long-baseline accelerator-generated beams from KEK~\cite{K2K:2002icj} and J-PARC~\cite{T2K:2011ypd}, while also searching for nucleon decay~\cite{Super-Kamiokande:2012zik,Super-Kamiokande:2014otb,Super-Kamiokande:2016exg,Super-Kamiokande:2020wjk}, dark matter~\cite{Super-Kamiokande:2004pou,Super-Kamiokande:2020sgt}, 
and both galactic~\cite{Super-Kamiokande:2007zsl,Super-Kamiokande:2016kji}  and diffuse supernova neutrinos~\cite{Super-Kamiokande:2021jaq,Super-Kamiokande:2002hei,Super-Kamiokande:2011lwo,Super-Kamiokande:2013ufi}.  The discovery of neutrino oscillations in SK's atmospheric neutrino 
data resulted in a share of the 2015 Nobel Prize in physics, while those results plus 
SK's solar and long-baseline neutrino measurements led to a share of {\em two} 2016 
Breakthrough Prizes.  As depicted in Figure~\ref{SK_hist}, Super-Kamiokande has operated 
under various configurations during its \mbox{long history}.  

Despite all of this success, one notable limitation SK had to operate under was the 
inability to efficiently detect thermal neutrons.  These were captured on free 
protons (hydrogen nuclei) in the pure water which filled Super-K, leading to the release 
of a single 2.2~MeV gamma.  Not only was this energy below typical SK trigger 
thresholds, but it 
also fell in an energy range strongly contaminated with backgrounds from a variety of
naturally occurring radioactive decays such as radon.  While great efforts have been 
made to overcome these limitations, the most advanced hydrogen-based studies still 
only achieved neutron tagging efficiencies around 20$\%$ at the cost of 1 in 100 of 
the copious accidental backgrounds getting through~\cite{Super-Kamiokande:2013ufi}.

\subsection{A Blend with Benefits}
To enable highly efficient neutron tagging while simultaneously providing  powerful background rejection, Beacom and Vagins first proposed a concept they 
called ``GADZOOKS!''  
(\underbar{G}adolinium \underbar{A}ntineutrino \underbar{D}etector \underbar{Z}ealously
\underbar{O}utperforming \underbar{O}ld \underbar{K}amiokande, 
\underbar{S}uper\underbar{!}), dissolving a 
gadolinium (Gd) salt -- such as gadolinium chloride, GdCl$_3$, or the somewhat less 
soluble but also considerably 
less corrosive gadolinium sulfate, Gd$_2$(SO$_4$)$_3$ -- in Super-Kamiokande's pure water~\cite{Beacom:2003nk}.  The primary goal of this proposal was to make observing the DSNB in 
Super-K possible; in fact, this paper is where the term ``DSNB'' was first introduced 
to help explicitly differentiate this subtle supernova neutrino signal from other 
``relic'' fluxes.

Gadolinium has the highest cross section for the capture of thermal 
neutrons of any naturally occurring stable substance, more than 100,000 times that 
of hydrogen, and following neutron capture the excited Gd nucleus emits an easily detected 
gamma cascade of $\sim$8~MeV.  This leads to a distinct IBD signature sometimes 
called the ``gadolinium heartbeat'': a prompt positron event followed a few 10s 
of microseconds later by a delayed neutron capture event.  The Cherenkov light of 
both events appears to originate nearly from the same place in the detector, as they 
typically occur close enough to fall within the position resolution of SK's vertex 
fitter.  Requiring such a double flash of light within such a short period of time, 
about 1/10$^{\mbox{th}}$ the delay for captures on hydrogen in pure water, serves 
to reduce accidental backgrounds by a factor of roughly 10,000, or 100 times cleaner 
than relying on captures on hydrogen alone.

\subsection{Putting the Gd in SK-Gd}
\begin{figure}[t]	
\centering
\includegraphics[width=0.67\textwidth]{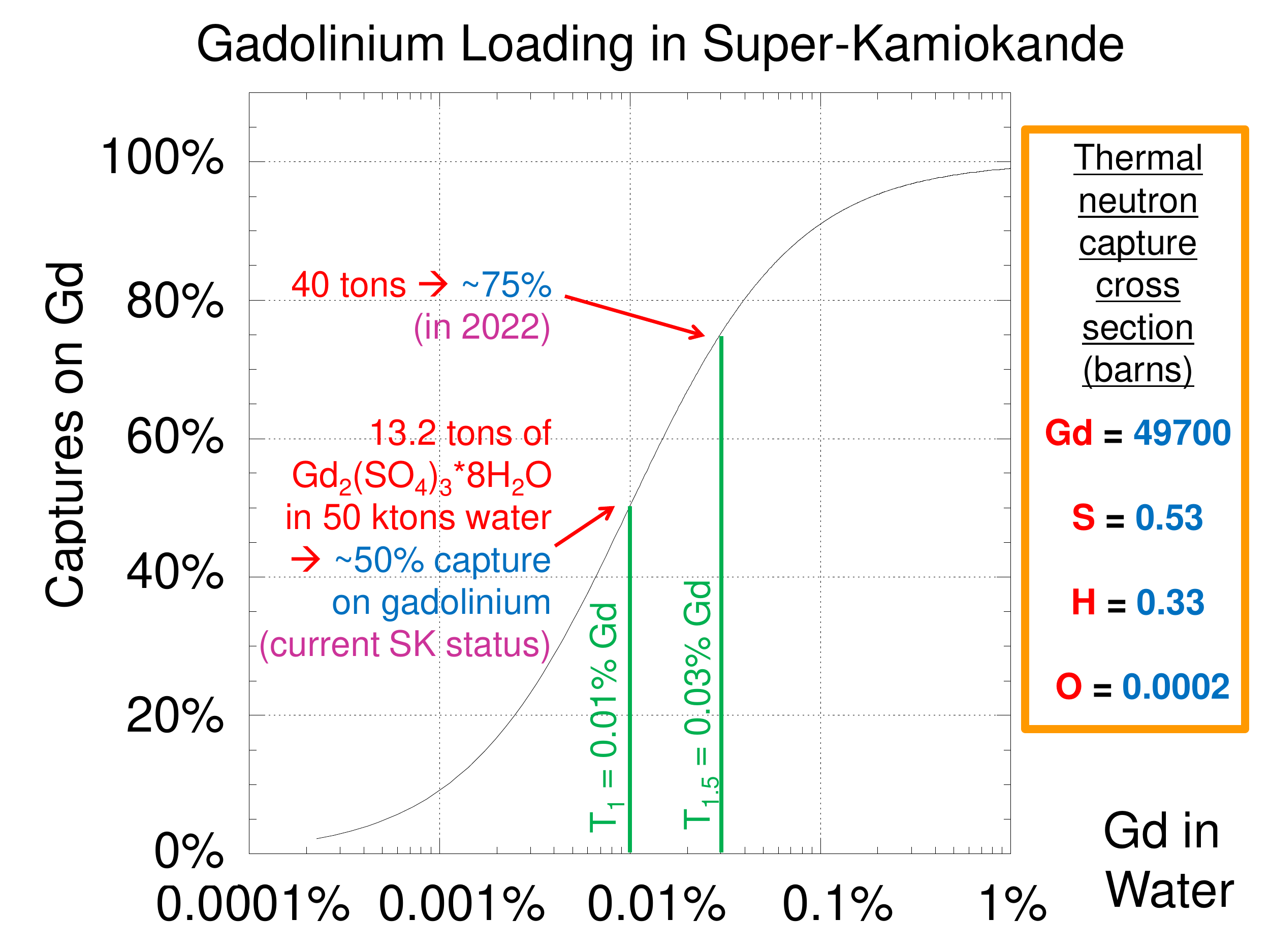}
\caption{Percentage of thermal neutron captures on gadolinium (Gd) as a function of 
dissolved mass percentage of Gd in water.  The first phase of loading in Super-Kamiokande 
is known as T$_1$, while the second phase is called T$_{1.5}$.  Thermal neutron capture 
cross sections of the four elements in the SK water are shown; nearly all neutrons 
not captured by Gd  end up on H as it is thousands of times more abundant 
inside SK than sulfur. \label{Gd_conc}}
\end{figure}

After years of R\&D to develop the necessary water filtration technology as
well as establish that loading gadolinium into Super-K would be both safe and effective~\cite{Marti:2019dof}, on 14 July 2020, the first 
dissolved gadolinium salt was injected into the SK detector.  This first stage of 
loading, which was completed on 17th August 2020, saw 13.2 tons of gadolinium 
sulfate octahydrate added to the SK water, resulting in a gadolinium concentration of
0.01$\%$ by mass~\cite{Super-Kamiokande:2021the}.  As shown in Figure~\ref{Gd_conc}, with 
0.01$\%$ Gd$^{3+}$ in 
solution about half of all thermal neutrons will visibly capture on the gadolinium, 
with the rest being collected near invisibly on hydrogen.  As everything has been 
running as expected, the Super-Kamiokande Collaboration plans to dissolve an additional 
27 tons of gadolinium sulfate octahydrate in 2022, bringing the total Gd ion concentration to 
0.03$\%$ by mass and the visible neutron fraction to 75$\%$.

It is expected that somewhere between 1 and 6 DSNB interactions with neutrino energies 
between 12 and 30~MeV should occur each year inside SK's fiducial volume of 22.5 ktons.
Assuming the middle of this range and taking into account detector efficiencies yields 
an expected DSNB signal rate of around 2.5 events per year with 0.03$\%$ Gd in  \linebreak the 
detector.  

As described above, there will be no remaining accidental backgrounds to speak of, and 
requiring the DSNB events to be above 12~MeV and below 30~MeV effectively suppresses the
physics backgrounds arising from nuclear power reactor antineutrinos causing low energy IBD
events and atmospheric neutrinos' charged current (CC) 
reactions, respectively.  Muon cuts in combination with the 12~MeV energy threshold will 
remove almost all background events 
caused by nuclear spallation, with the efficient neutron tagging now provided by gadolinium
allowing even better spallation cut efficiencies than those employed by SK to 
date~\cite{Super-Kamiokande:2011lwo,Li:2015kpa,Li:2015lxa}. Most of the remaining physics background are therefore 
expected to come from neutral current (NC) interactions involving energetic 
atmospheric neutrinos interacting with oxygen nuclei, but a recent paper has 
shown that these can be significantly and efficiently suppressed through the 
use of a machine learning (specifically a convolutional neural network) approach, 
removing 98$\%$ of the NC background at the expense of just 4$\%$ of the signal yielding 
a signal-to-background rate of 4:1~\cite{Maksimovic:2021dmz}.  In all, we rather conservatively  
assume a total residual background rate of 0.8 events per year in this energy range.

\subsection{The Future of Gd-Loaded Water Cherenkov Detectors}
\label{sec:hk}

Data collection in the Gd-enhanced Super-Kamiokande has been underway since the middle 
of 2020, and is expected to continue until at least 2028.  In 2027, the 
new Hyper-Kamiokande (Hyper-K, HK) detector, some eight times the fiducial 
volume of SK and currently under construction, is scheduled to come online~\cite{Abe:2011ts,Hyper-Kamiokande:2021frf}.  As was the case with Kamiokande ceding the field to Super-Kamiokande 
and turning off in 1997, it is expected that Super-K will also be permanently decommissioned 
once Hyper-K is complete and operating stably.  While HK will not contain gadolinium 
on Day 1, it is assumed that gadolinium will very likely be added to the new 
detector eventually, such that all proposed HK detector components and materials 
must be certified to be compatible with extended immersion in Gd-loaded water.  
From simple scaling, a Gd-loaded Hyper-K can be expected to observe an SN1987A-like 
number of supernova neutrino events from the DSNB every year it is in operation, an 
exciting prospect indeed.

\section{The Juno Experiment}
\label{sec:juno}





\begin{figure}[t]	

\includegraphics[width=1\textwidth]{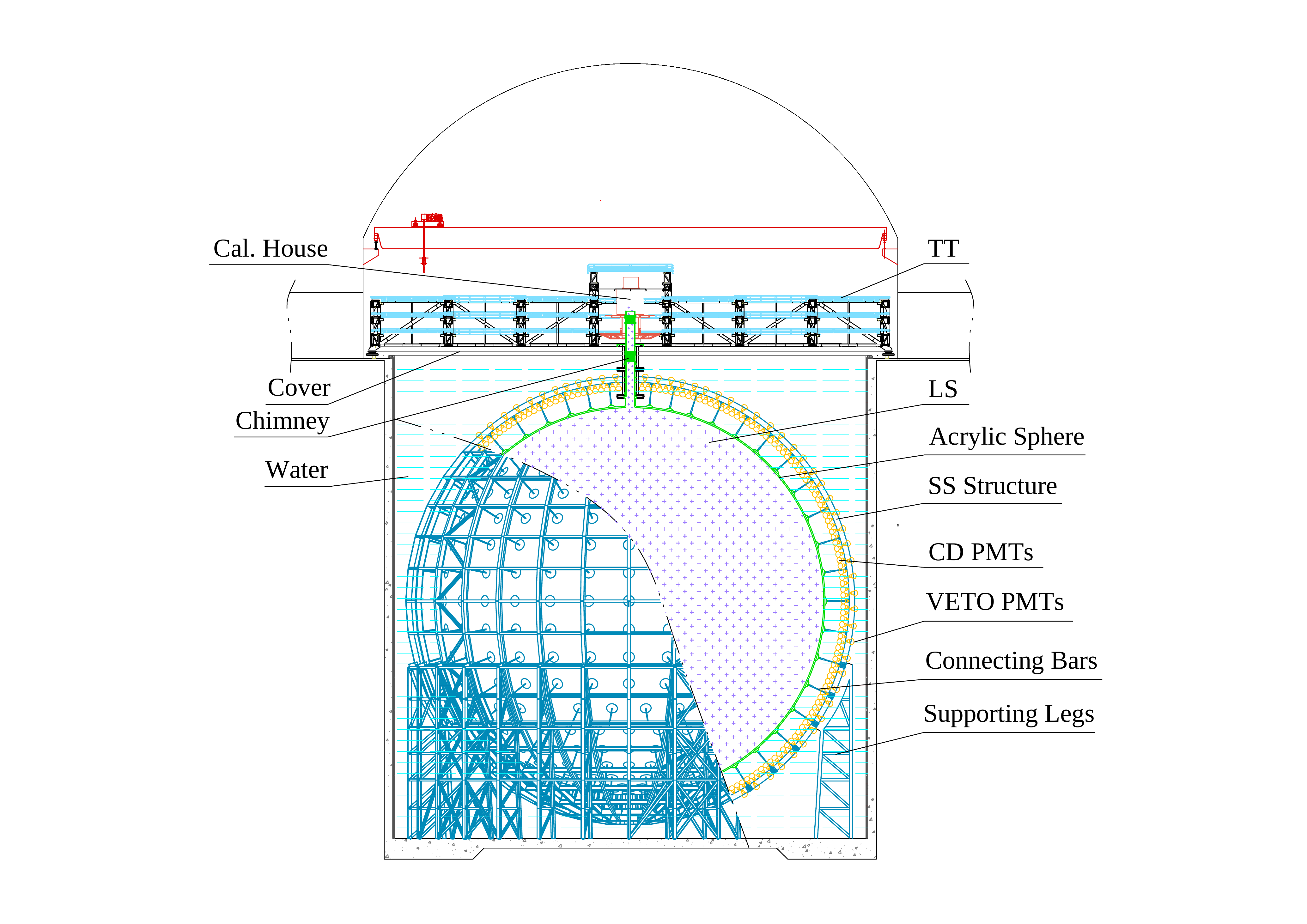}
 \caption{An illustration for the JUNO detector detector complex. The figure is taken from {Ref.} 
~\cite{JUNO:2021vlw}. \label{JUNO_detector}}
\end{figure}

The JUNO experiment is located at Jiangmen in South China. Its primary goal is to determine the neutrino mass ordering and precision measurements of neutrino oscillation parameters using reactor neutrinos from the powerful Taishan and Yangjiang nuclear power plants~\cite{JUNO:2015zny,JUNO:2021vlw}.
JUNO will build a Liquid Scintillation (LS) detector of 20 kton with an overburden of 700 m rock for shielding the cosmic rays.
As a multiple-purpose neutrino observatory, the JUNO detector complexes, from the inner to outer layers, include the Central Detector (CD), the Veto Detectors and the Calibration System. An illustration for the JUNO detector detector complex is provided in Figure~\ref{JUNO_detector}.
The CD contains 20 kton LS in an acrylic shell with an inner diameter of 35.4 m, and 17,612 high-quantum-efficiency 20-inch Photo Multiplier Tubes (PMTs) and around 25,600 3-inch PMTs are closely packed around the LS sphere in order to guarantee the precision neutrino energy measurement with the energy resolution of 3\%~\cite{JUNO:2020xtj}.
Other sub-systems include the water pool and top tracker veto system, the calibration system, the online LS monitoring system, and a satellite TAO reference reactor spectrum detector~\cite{JUNO:2020ijm}.  
JUNO is expected to take data in 2023.

The primary detection channel for the DSNB is the IBD reaction on free protons, in which the prompt positron signal takes away most of the neutrino energy, and the delayed neutron capture signal is correlated with the prompt signal with distinct energy, time interval, and spatial interval relations.
With the different model predictions mentioned in Section~\ref{sec:dsnb1}, it is estimated that around 1--5 DSNB IBD events per year can be observed between 12 and 30 MeV~\cite{JUNO:2021vlw}. After background cuts, there remain 1.4 IBD events per year for the DSNB \emph{reference model} (see below).

Compared to water Cherenkov detectors, LS detectors such as JUNO have intrinsically high neutron tagging efficiencies for neutron capture on free protons. Given the high scintillation light yield, the 2.2\,MeV gamma rays emitted in the capture provide a delayed signal easy to identify. Given the excellent vertex reconstruction capabilities and expected low background levels, prompt and delayed signals can be correlated with high efficiency, close to unity in the LS bulk volume.

In the visible energy of interest relevant to the DSNB search, there are different categories of backgrounds in JUNO:

\begin{itemize}
\item First, there are two intrinsic backgrounds from other $\overline{\nu}^{}_e$ sources. In the vicinity of the low energy part of the DSNB $\overline{\nu}^{}_e$ spectrum, the irreducible background is from those $\overline{\nu}^{}_e$'s emitted from nearby nuclear power reactors, whose fluxes are highly decreased above the neutrino energy of around $\mathcal{O}(10)$ MeV. A choice of the lower boundary of the search window at 12 MeV can reduce this background to a negligible level. 
The high energy part of the indistinguishable background is composed of the IBD interactions of the low energy tail of atmospheric $\overline{\nu}^{}_e$ with free protons.

\item The second category of the main backgrounds for the DSNB searches is from the cosmic muon spallation. 
Fast neutrons are generated by spallation events outside the CD. The event rate is higher for larger radii, in particular within the upper and equator regions because of the shallow water buffer. Therefore, the fast neutron background can be reduced by proper selection of the fiducial volume of the CD.
The $^{9}$Li/$^{8}$He background is produced from radioactive decays of long-lived spallation isotopes in the CD, and is correlated with the parent muons and associated neutrons. Therefore, the $^{9}$Li/$^{8}$He background can be effectively reduced by muon veto strategies. Moreover, excellent energy resolution at JUNO will ensure most of the $^{9}$Li/$^{8}$He background below 12 MeV of the visible energy and can be safely neglected if 12 MeV is chosen as the lower boundary of the search window.

\item Finally the dominant background for the DSNB search is from the neutral current (NC) interactions of atmospheric neutrinos with the carbon nuclei. When high energy atmospheric neutrinos interact with carbon, copious neutrons, protons, $\gamma$'s and $\alpha$'s are generated in association with the outgoing leptons, where those interaction channels with single neutron production may contaminate the IBD signals.
To model the NC interaction between the atmospheric neutrinos and the carbon nuclei, one needs to employ both the neutrino interaction generator tools~\cite{Andreopoulos:2009rq,Golan:2012rfa} 
and the package for deexcitations of the final-state nuclei~\cite{Koning:2005ezu}. A careful investigation of the atmospheric neutrino NC background has been accomplished in 
Refs~\cite{Cheng:2020oko, Cheng:2020aaw}, which are shown to be larger than the DSNB signal by one order of magnitude.
\end{itemize}

Pulse shape discrimination (PSD) is expected to be a very efficient technique to further improve the signal-to-background ratio. Regarding all the possible IBD-like backgrounds, the prompt signal of fast neutron and atmospheric neutrino NC events is predominantly created by heavy particles such as neutrons, protons and $\alpha$'s. In LS detectors, the distinct time profiles of different types of particles permit effectively distinguishing between the light $\gamma$-like particles (i.e., $e^{+}$, $e^{-}$, and $\gamma$) and heavy proton-like particles (i.e., proton, neutron, and $\alpha$). By virtue of the high light yield and excellent time resolution at JUNO, it is estimated that the atmospheric NC background 
can be reduced by two orders of magnitude while the signal efficiency of the DSNB remains at least above 50\%~\cite{JUNO:2021vlw}. 
Recent studies indicate that JUNO's sensitivity could be substantially improved based on a refined scheme for the PSD-based particle identification~\cite{Cheng:2021ais}.


To summarize, with all the possible background contributions and suppression techniques are taken into account, a total background level of 0.7 events per year is estimated and---depending on the DSNB event rate---an excellent signal-to-background ratio of 1:1 to 4:1 can be achieved. Therefore, we can anticipate a good discovery sensitivity of the DSNB in the coming decade. 


\section{Projected Dsnb Sensitivities}
\label{sec:sensitivity}

\begin{figure}[t!]	
\includegraphics[width=0.75\textwidth]{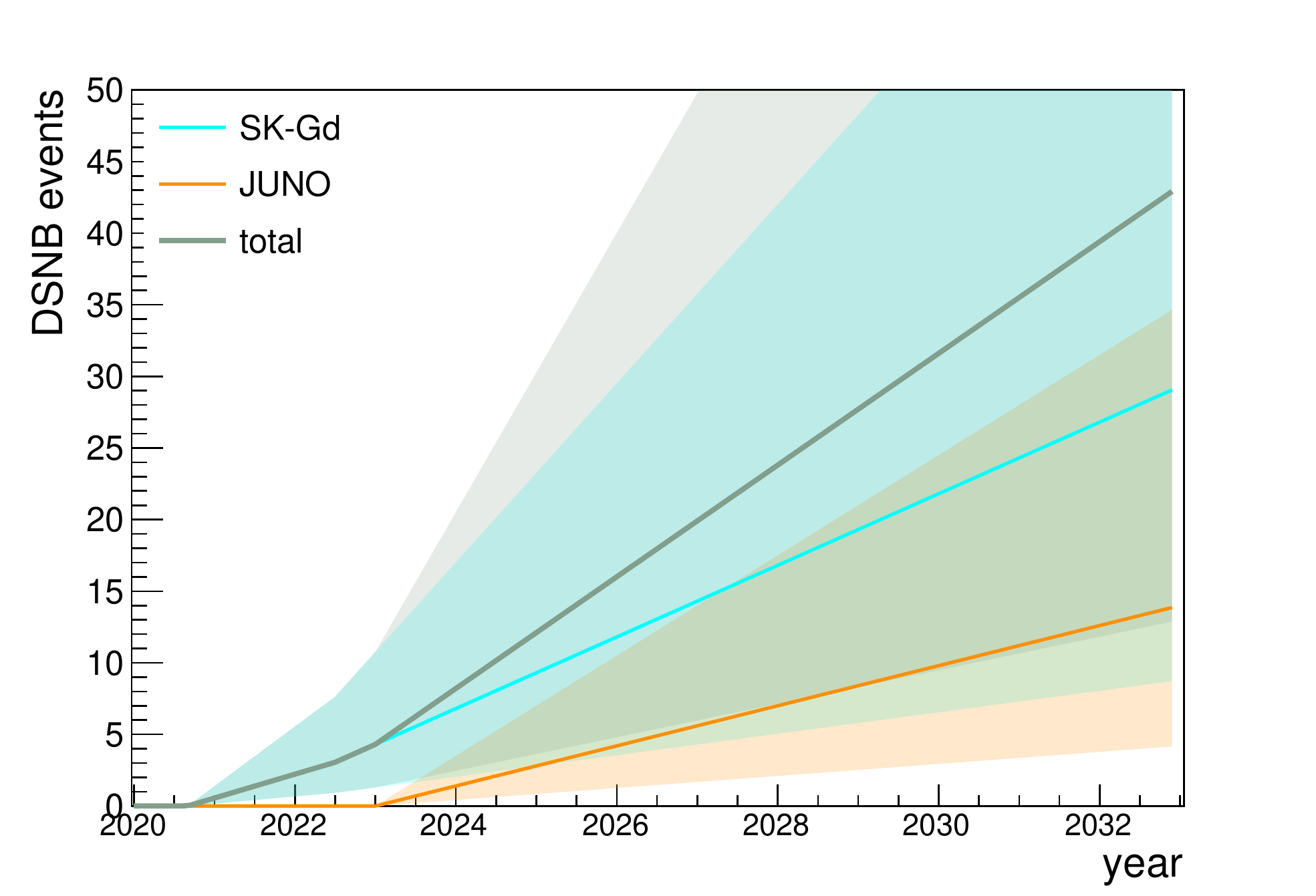}
\caption{Time development of the cumulated number of signal event for SK-Gd and JUNO, using the time and rate information as quoted in Table~\ref{tab:rates}. The solid lines correspond to the rates of the DSNB \emph{reference model}, the shaded areas reflect the range implied by the variability of the signal (Section~\ref{sec:dsnb1}). Please note that recent studies for JUNO indicate that a higher signal efficiency could be reached~\cite{Cheng:2021ais}, thus the event rates per year for both experiments are rather compatible.}
\label{fig:rates}
\end{figure} 

Even in experiments the size of Super-Kamiokande and JUNO, accumulating the data for a DSNB detection is a waiting game. Table \ref{tab:rates} provides a short summary of the DSNB signal efficiences and detected event rates as well as the background rates expected for SK-Gd and JUNO for the DSNB \emph{reference model} described in Section \ref{sec:dsnb1}. The following assumptions are made in the calculations:
\begin{itemize}
\item We refer to the nominal fiducial masses, i.e., 22.5\,kt (or $N_p=1.50\times10^{33} $) for SK-Gd and 17\,kt ($N_p=1.22\times10^{33}$) for JUNO. 

\item For easy comparison, we choose in both cases the same observation window, ranging in visible energy from 12\,MeV to 30\,MeV. This range is defined by the irreducible backgrounds for the DSNB observation, i.e., reactor and atmospheric $\bar\nu_e$ fluxes. Please note that while the reactor $\bar\nu_e$ background at the location of SK will be smaller, this advantage is at least partially compensated for by the better energy resolution of JUNO \cite{JUNO:2021vlw}. 

\item For background rates, we use the numbers lined out in sections \ref{sec:sk} and \ref{sec:juno}. The dominant contribution in JUNO is formed by NC interactions of atmospheric neutrinos. In SK-Gd, invisible muons will play an important role.

\item Finally, for SK-Gd, we cite two sets of numbers in dependence of the gadolinium concentration that is set to be increased in mid-2022 from 0.01\% to 0.03\% (Section~\ref{sec:sk}). 
\end{itemize}
Please note that both experiments feature a rather similar ratio of signal ($S$) and background ($B$) rates of $S:B\sim2$. 
\begin{table}[t] 
\setlength{\tabcolsep}{2.5mm}

\caption{DSNB fiducial masses, signal efficiencies and rates as well as background rates expected for SK-Gd and JUNO~\cite{JUNO:2021vlw} for the DSNB \emph{reference model} in the energy range of 12$-$30\,MeV. The two rows quoted for SK-Gd reflect the conditions for the initial 0.01\% as well as the increase to 0.03\% Gd loading foreseen for mid-2022. \label{tab:rates}}
\centering
\begin{tabular}{cccccc}
\hline
	& \textbf{Fiducial} &  	&  \textbf{Signal}	& \textbf{Signal}	& \textbf{Background} \\
\textbf{Experiment}	& \textbf{Mass} \textbf{[kt] }& \textbf{Time Range} & \textbf{Efficiency} & \textbf{Rate} \textbf{[yr}$^{\boldmath-1}$\textbf{]} & \textbf{Rate} \textbf{[yr}$^{\boldmath-1}$\textbf{]}\\
\hline
SK-Gd	&   22.5 & 8/20--06/22 & 50\%  &  1.7  &  0.8   \\
        &        & 7/22--      & 75\%  &  2.5  &  1.2   \\
\hline
JUNO	&   17.0 & 1/23--     & 50\%  &  1.4  &  0.7  \\
\hline
\end{tabular}
\end{table}

The discovery potential for the DSNB lastly depends on the total number of signal and background events accumulated over a longer period of measuring time. Figure \ref{fig:rates} displays the time development of the DSNB signal rates over time. For this, we used the information given in Table~\ref{tab:rates} regarding signal and background rates as well as the different dates for start of data taking (and SK-Gd upgrade). We show as well the development of the total number of DSNB events. A level of $\sim$40 DSNB events detected is reached after 10 years. This number refers to our \textit{DSNB reference model} (Section~\ref{sec:dsnb1}). Based on the uncertainties of the DSNB signal prediction, the actual event number and thus rate of signal collection might substantially deviate from the reference prediction. 
The corresponding ambiguity is reflected by the shaded areas. Naturally, a low signal rate would affect both experiments in the same way. Therefore, the shaded regions should not be mistaken to be classical uncertainty bands but are instead fully correlated.

Given the earlier start, larger fiducial mass and higher efficiency after the increase in Gd concentration, SK-Gd is expected to accumulate statistics somewhat faster than JUNO. However, we note here that recent studies for JUNO indicate that a higher signal efficiency could be reached using a more advanced method of pulse shape discrimination, bringing both experiments roughly on par~\cite{Cheng:2021ais}.

Based on these numbers, it becomes possible to estimate the experimental sensitivities of the individual and combined measurements. While the eventual DSNB analyses will apply more sophisticated techniques, here we restrict ourselves to a simple count rate analysis for signal and background in the energy window of interest (12$-$30\,MeV). As a measure of sensitivity, we adopt the ratio $S/\sqrt{S+B}$, i.e.,~the significance of the signal strength over the expected statistical variation of the count rate. Clearly, this simplified approach has many short-comings. Most notably, it neglects the relevant systematic uncertainties in the predicted background rates. However, it provides an easy-to-understand measure of the sensitivity, its development over time and permits the comparison and combination of \mbox{the experiments}.

\begin{figure}[t]	
\includegraphics[width=0.75\textwidth]{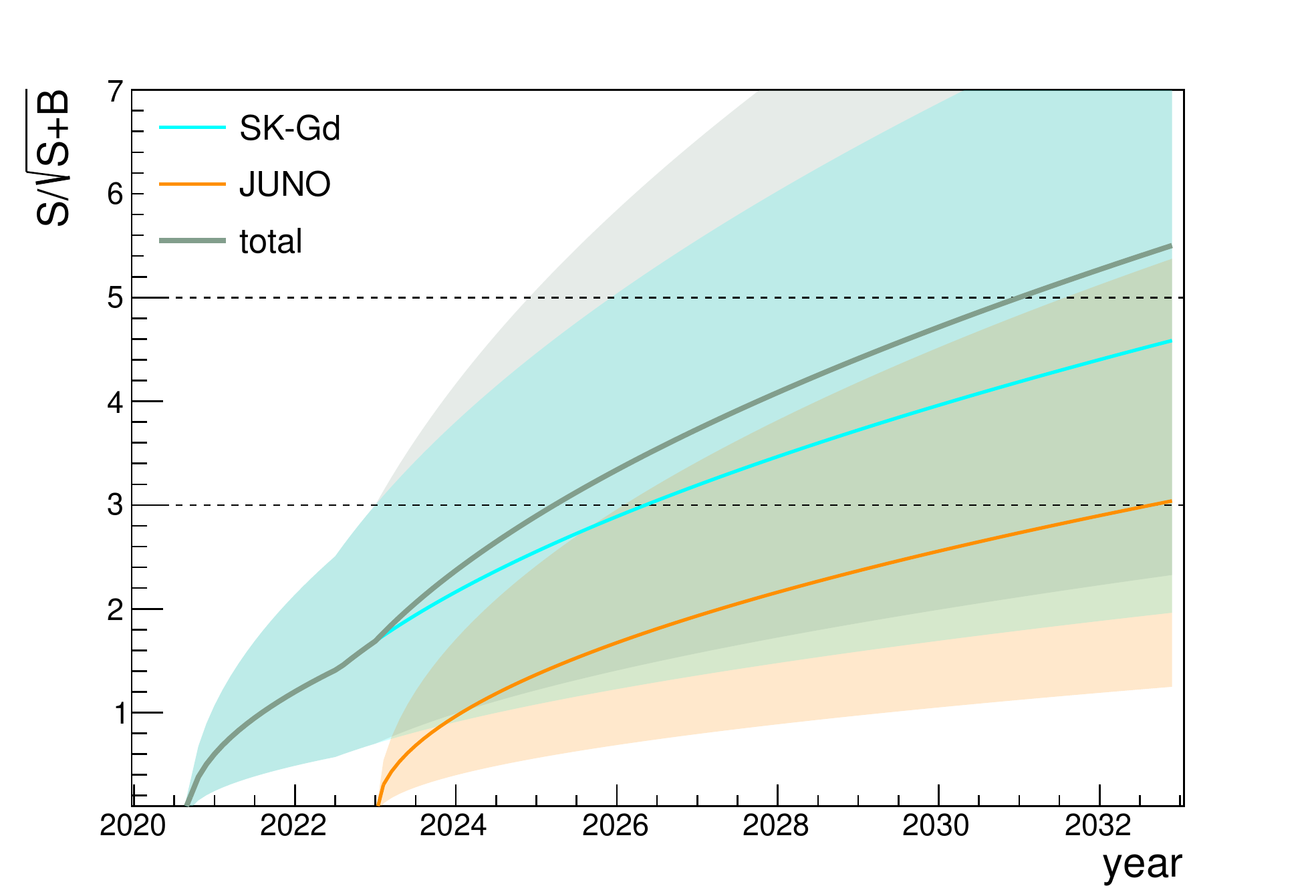}
\caption{Statistical significance of a DSNB signal rate $S$ excess over background rate $B$ based on the figure of merit $S/\sqrt{S+B}$. Individually, both experiments can reach a $3\sigma$ significance level over about 10 years of measuring. The combined sensitivity reaches $5\sigma$ in the early 2030s.
\label{fig:sensitivity}}
\end{figure}

Using the signal and background numbers listed in Table \ref{tab:rates}, we display the time-development of the signal significance in Figure \ref{fig:sensitivity}. We display the significance levels for both experiments individually and for their combination. While the solid lines correspond to the DSNB \emph{reference model}, the shaded areas indicate the predicted signal range. Both experiments individually are expected to reach a $3\sigma$ statistical evidence of the DSNB signal within a decade of measuring time, with a clear lead of SK-Gd due to the earlier start of measurement and faster signal accumulation.




The combined sensitivity curve of Figure \ref{fig:sensitivity} illustrates that the sum signal of both experiments could be used to achieve a level of $5\sigma$ observation of the DSNB \emph{reference model} within the next 10 years. The corresponding signal and background rates as well as statistical sensitivities are summarized in Table \ref{tab:comparison}. As before, we have neglected systematic uncertainties on the estimated background levels. Arguably, a combined analysis might achieve better sensitivity, since both experiments will collect somewhat complementary data sets on the atmospheric neutrino NC background (Cherenkov vs.~scintillation signals) that are potentially useful to better constrain the associated systematic uncertainties on background rate and spectrum.

Finally, it should be noted that---even if the data sets of both experiments were combined---only several tens of signal events are expected for the \emph{reference model}, reaching close to $10^2$ under the most optimistic assumptions. Consequently, the spectral information that can be obtained from this next generation of DSNB experiments will be rather limited, at best comparable to the accuracy gained from the neutrino burst of SN1987A. Therefore, while indeed a first positive detection of the DSNB is within reach within the next decade, a substantially larger detector such as HK-Gd (i.e.,~with enhanced neutron tagging) will be required to extract details on the DSNB spectrum, thus offering a window to the underlying physics of SN core collapse, black-hole formation and redshift-dependent collapsar rate.

\begin{table}[t] 
\setlength{\tabcolsep}{5.2mm}

\caption{Cumulated number of signal ($S$) and background ($B$) events expected for SK-Gd and JUNO for mid-2031, i.e., about 11 years after the start of the SK-Gd measurement. The quoted figure of merit $S/\sqrt{S+B}$ corresponds roughly to the signal sensitivity in standard deviations. The combined sensitivity reaches $5\sigma$ at this time. \label{tab:comparison}}

\begin{tabular}{ccccc}
\hline
	&   \textbf{Mesuring}	&  \textbf{Signal} & \textbf{Background} & \textbf{Sensitivity}\\
\textbf{Experiment} &  \textbf{Time} \textbf{[yrs]} & \textbf{($\boldmath S$)}	& \textbf{($\boldmath B$)} & \textbf{($\boldmath S/\sqrt{\textbf{\emph{S}}+B}$)}\\
\hline
SK-Gd	&   11 &  26  &  12 &  4.1    \\
JUNO	&  8.5 &  12   & 6    & 2.7         \\
\hline
total	&    & 38   & 18   &   5.0           \\
\hline
\end{tabular}
\end{table}

\section{Conclusions}

The start of SK-Gd data taking in late 2020 and the expected start of JUNO data taking in 2023 indicate a substantial improvement of the worldwide sensitivity for diffuse Supernova neutrinos (or, more precisely, its $\bar\nu_e$ component). Given the large unknowns of the signal flux and spectrum and the potential systematics associated with background rates and subtraction, it is difficult to forecast the exact level of sensitivity to be achieved by the two experiments. However, using our DSNB \emph{reference model} (Section \ref{sec:dsnb1}) and making simplified assumptions on the signal significance (Section \ref{sec:sensitivity}), we can conclude that both experiments on their own are likely to obtain statistical evidence of the signal  ($3\sigma$ level) within about 10 years of running time. The combination of their results may even allow a $5\sigma$ discovery of the DSNB in the same time frame. After more than 20 years of experimental searches, a first observation of the DSNB signal seems thus well in reach within the \linebreak next decade.

\section*{Acknowledgments}
YFL' work was supported in part by National Key R\&D Program of China under Grant No. 2018YFA0404101, by National Natural Science Foundation of China under Grant Nos. 11835013, and by the CAS Center for Excellence in Particle Physics.
M.V.'s work was supported by MEXT KAKENHI Grant Numbers 17H06357 and 17H06365.
M.W.'s work was supported by the DFG Research Unit ''JUNO'' (FOR 2319).

\bibliographystyle{apsrev4-1}
\bibliography{arxiv_version_v2.bib}

\end{document}